# Controlling the rate of resonant tunneling between quantum wells by relocating the wave function within a quantum well.

Yu.A.Mityagin, M.P.Telenkov

P.N.Lebedev Physical Institute of Russian Academy of Sciences,

119991 Leninsky pr., 53, Moscow, Russia

The possibility of controlling the tunneling time between quantum wells by relocation of the subband wave function within the quantum well by varying the configuration of thin tunnel-transparent barriers embedded into the well is demonstrated.

When considering resonance-tunnel transport in superlattices and quantum well structures in the regime of sequential tunneling, it is usually assumed that the rate (time) of tunneling is determined primarily by the width and height of barriers separating the wells. Accordingly, to change the tunneling rate it is necessary to vary the barrier widths.

In this work, we demonstrate a different way to control the amplitude of tunneling resonance, which consists of redistributing the wave function within the quantum well, either bringing the maximum of its distribution close to the barrier or moving away from it. In the first case, the degree of penetration of the wave function into the barrier will increase, leading to an increase in the probability of tunneling and reducing tunneling time, in the second case - it will decrease, reducing the probability of tunneling.

To relocate the wave function within the quantum well, we consider introducing a series of thin tunnel-transparent barriers into the well, as previously proposed to regulate the energy of the lower subband [1-3]. Relocation of the wave function in this case is provided by varying the position of the barriers inside the well while keeping the energy of the lower subband unchanged.

We will consider the tunnel resonance between two $GaAs/Al_{0.3}Ga_{0.7}As$ quantum wells of fixed width $d_w$=25 nm separated by a barrier of width $d_b$. In each of the wells, we introduce three barriers of 2.5 nm width each.

The energies and wave functions of the electrons were determined by solving the Schrödinger equation for the envelope in the parabolic approximation.

As an initial situation, we take the case when all the distances between the barriers and the pit walls are the same (Fig.1a). In this case, the wave function of the electron in each of the quantum wells is symmetric and penetrates the barriers to the right and left of the well in the same way.

When the position of the barriers in the well is changed, namely, when they are shifted toward one of the edges, the maximum of the wave function distribution shifts to the opposite edge of the well, as shown in Fig. 1b. In this case, the degree of penetration of the wave function into the barrier increases, which should lead to an increase in the tunneling probability.

To illustrate the effect, we considered several configurations of the three barriers introduced into the quantum well, shown in Table 1. The initial configuration - symmetric (number 0), is shown in Fig. 1.

For all configurations, the barriers in the well were selected so that the energy of the lower subband (70 meV) remained unchanged.

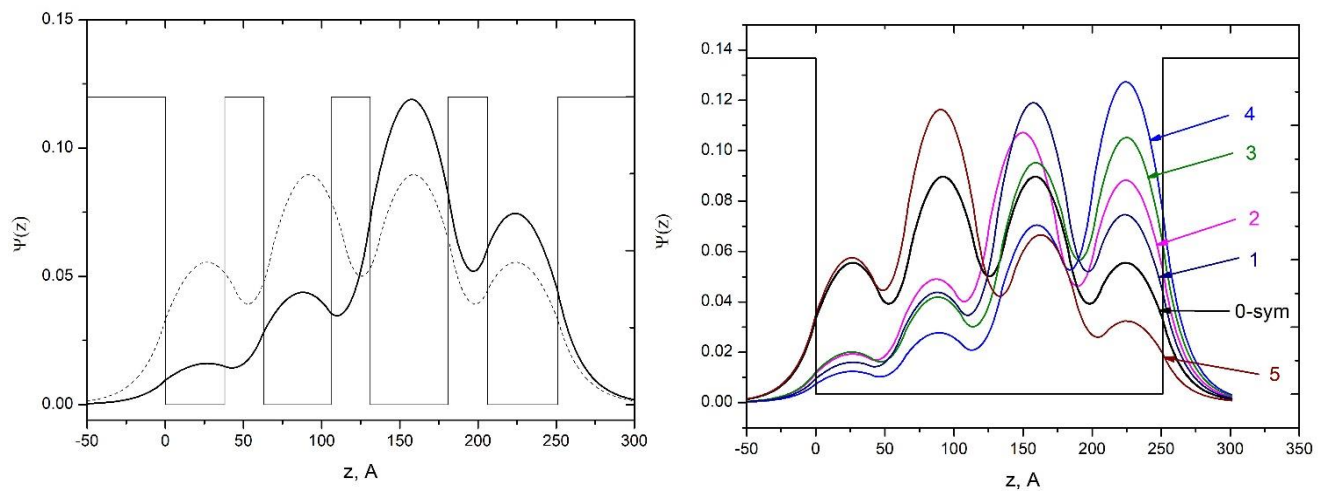

Fig.1. a). Wave functions of the 1st subband in a quantum well with 3 introduced barriers - for symmetric barrier configuration (dashed line) and for barrier configuration number 2 from Table1 (solid curve). Well potential profile shown corresponds to barrier configuration number 2.

(b) Transformation of the wave function of the 1st subband in a quantum well with 3 introduced barriers when the barrier configuration is changed. The numbers of curves correspond to configurations listed in Table 1.

Let us now consider two identical quantum wells of width 25 nm separated by a barrier of width db. In each of the wells, we will introduce three thin (2.5 nm width) barriers, varying their location to shift the maximum of the wave function distribution, either bringing it closer to the separating barrier (Fig.2a) or away from it (Fig.2b), according to the configurations given in Table 1. Accordingly, the barrier configuration of the second well is mirror symmetric to the one in the first well. In this case the level of the lower subband is split by the value $\Delta E_1$, determined by the value of the integral of the overlap of wave functions of each of the wells. Naturally, this value is also determined by the width $d_b$ of the barrier separating the wells.

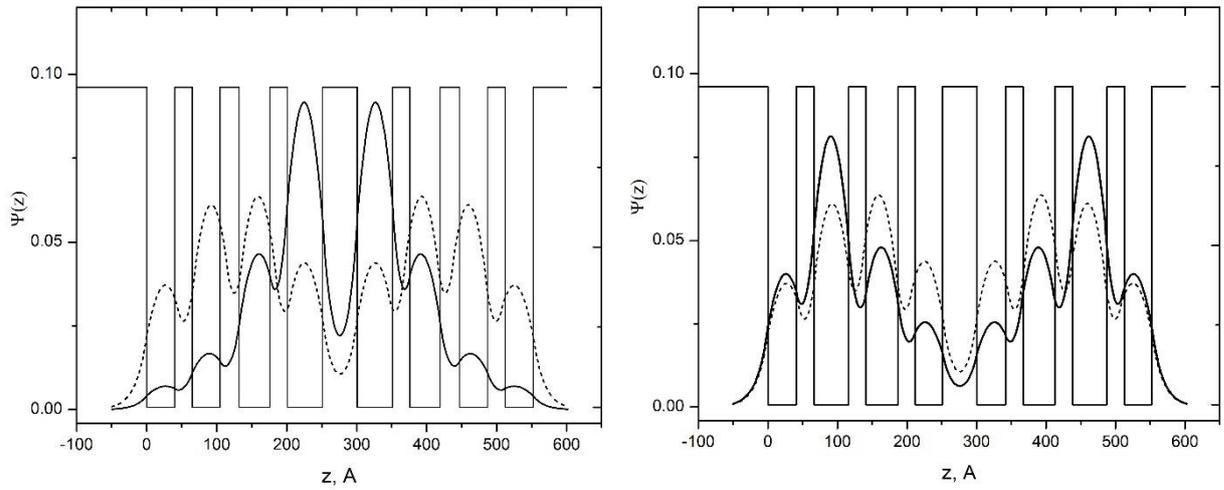

Figure 2. Wave functions of the 1st subband of two quantum wells with 3 introduced barriers each – (a) for configuration number 4 and (b) for configuration number 5 from Table 1. The wave functions for the symmetric barrier configuration are shown by the dashed line in both figures.

|         | w  | b  | w  | b  | w  | b  | w  |
|---------|----|----|----|----|----|----|----|
| 0(symm) | 44 | 25 | 44 | 25 | 44 | 25 | 44 |
| 1       | 38 | 25 | 40 | 25 | 49 | 25 | 45 |
| 2       | 38 | 25 | 38 | 25 | 51 | 25 | 49 |
| 3       | 39 | 25 | 40 | 25 | 48 | 25 | 49 |
| 4       | 40 | 25 | 40 | 25 | 46 | 25 | 50 |
| 5       | 41 | 25 | 50 | 25 | 46 | 25 | 39 |

Table 1. Configurations of barriers introduced into the quantum well. The widths of barriers and spaces between them are given in angstroms.

Fig. 3a shows the calculated dependences of $\Delta E_1$ on the barrier width $d_b$ for the configurations of the embedded barriers from Table 1. Configurations 1-4 correspond to the gradual approach of the maximum of the wave function distribution to the barrier, and configuration 5 - to the shift away from the barrier. It can be seen that by varying the position of the barriers it is possible (within a sufficiently wide range) to both increase and decrease the value of $\Delta E_1$ at each fixed value of $d_b$. To estimate the tunneling time, we will use the well-known approach [4], which defines the tunneling time as half of the period of Rabi oscillations:

$$t_{tunn} = \frac{\pi \hbar}{\Delta E_1}.$$

The tunneling times thus obtained as a function of the barrier width $d_b$ are shown in Fig. 3b. The figure shows that by changing the configuration of the barriers and thereby relocating the wave function within each of the quantum wells, it is possible to vary the tunneling time between the wells - increasing or decreasing - by more than an order of magnitude.

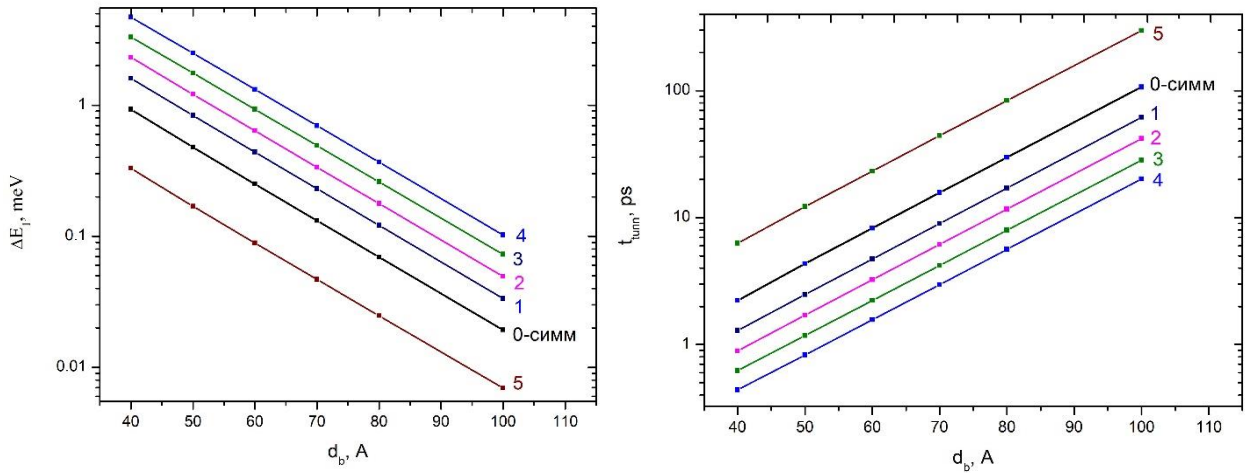

Fig.3. Dependences of the lower subband splitting energy (a) and inter-well tunneling time (b) on the width $d_b$ of the barrier separating the wells for two neighboring quantum wells with 3 embedded barriers in each. The data are shown for the configurations of the barriers presented in Table 1.

It should also be noted that the wave function can be similarly relocated simultaneously to the right and left edges of the well by shifting the barriers symmetrically to the center of the well. In this case, the degree of wave function penetration into the barrier on the right and left sides will also increase (or fall if the embedded barriers are shifted from the center of the well to its edges). This approach can be used when developing periodic resonance-tunneling structures from quantum wells, for example, superlattices.

Thus, the possibility of controlling the tunneling time between quantum wells by relocating the wave function of a subband within a quantum well has been demonstrated. The relocation can be accomplished by embedding a series of thin tunneling-transparent barriers into the quantum well and varying barrier positions within the well. It is possible to vary the inter-well tunneling time by more than an order of magnitude without changing the width of the barrier separating the wells. I.e. it is possible to achieve rather small tunneling times at relatively wide barriers. This technique can be used in designing quantum well resonant-tunnel structures such as resonant-tunnel diodes, quantum cascade lasers, superlattices, in which the widths of minibands can be varied in this way, etc.